# Electronic Transport Properties of Carrier Controlled SnSe Single Crystals

Aichi Yamashita[1,2,‡], Osamu Ogiso[1,2], Ryo Matsumoto[1,2], Masashi Tanaka[1,3],

Hiroshi Hara[1,2], Hiromi Tanaka[4], Hiroyuki Takeya[1], Chul-Ho Lee[5], and

Yoshihiko Takano[1,2]

[1] WPI-MANA, National Institute for Materials Science, 1-2-1 Sengen, Tsukuba, Ibaraki,

305-0047, Japan

[2] Graduate School of Pure and Applied Sciences, University of Tsukuba, 1-1-4 Tennodai,

Tsukuba, Ibaraki, 305-8577, Japan

[3] Graduate School of Engineering, Kyushu Institute of Technology, 1-1 Sensui-cho,

Tobata-ku, Kitakyushu-shi, Fukuoka, 804-8550, Japan

[4] National Institute of Technology, Yonago College, 4448 Hikona, Yonago, Tottori

683-8502, Japan

[5] National Institute of Advanced Industrial Science and Technology, Tsukuba, Ibaraki,

305-8568, Japan

We found that the electronic transport property of SnSe single crystals was sensitive to oxygen content. Semiconducting SnSe single crystals were obtained by using Sn of grain form as a starting material while powder Sn resulted in metallic SnSe. X-ray photoelectron spectroscopy analysis revealed that the surfaces of raw Sn were oxidized, where the volume fraction was relatively low in grain Sn. This demonstrates that contamination of oxygen causes metallic behavior in grown SnSe single crystals.

Single crystalline SnSe has been recently shown to have an ultrahigh ZT value of 2.6 and 2.3 at 923 K in the *b* and *c* directions, respectively [1]. These high ZT values are achieved with high power factor and extremely low lattice thermal conductivity. After the report by Zhao *et al*. in 2014, intense investigations have been conducted by many groups [2-19]. However, the reproducibility of this exceptionally high thermoelectric performance has not been obtained to date. Different thermoelectric properties, especially in electrical conductivity, have been reported from several groups for both of single crystalline and poly crystalline samples. Some groups reported metallic behavior in electrical conductivity [12-14]. On the other hand, others reported semiconducting behavior [15-19] and consistent results have not been obtained so far.

Motivated to figure out the reason of these different results, we focused on the oxygen contamination from the starting materials of Sn due to the difference of oxidized surface area. Herein, we report on the electrical conductivity measurements along the *bc* plane for SnSe single crystals in a temperature range from 2 to 390 K. It was found that the electrical conductivity was controlled by changing the oxygen amount during the synthesis. The electrical conducting behavior changed from semiconductor to metal with increase of the oxygen amount.

Single crystals of SnSe were grown from the congruent melt. Starting materials were

Se chips and two different forms of Sn. Grain form Sn (diameter: ~ 3.0 mm) and powder form Sn (diameter: ~ 0.15 mm) with the same purity of 99.99 % were used. To examine the form dependence on electric transport property of grown single crystals, the ratio of grain and powder form Sn was systematically varied (grain Sn : powder Sn = 100 : 0, 90 : 10, 70 : 30, 0 : 100). These Sn grains and/or Sn powders and Se chips were loaded into quartz tubes in stoichiometric composition. The tubes were then evacuated and sealed. They were slowly heated to 600$^o$C and kept for 12 hours followed by furnace cooling. As a result, poly crystalline ingot of SnSe was obtained. The obtained ingot was ground and heated to 1000$^o$C and kept for 10 hours in the double sealed evacuated quartz tubes. These double quartz tubes are necessary to avoid the enters of the oxygen by the breaks of inner quartz tube due to stress from the structural transition during furnace cooling. They were cooled down to 840$^o$C in 8 hours followed by furnace cooling. Finally, SnSe single crystals with typical size of 10 × 5 × 3 mm$^3$ were obtained.

Powder X-ray diffraction (XRD) patterns were measured by using Mini Flex 600 (Rigaku) with Cu-Kα radiation. For the measurements, single crystals were ground into powder. Chemical states of raw Sn were examined by X-ray photoelectron spectroscopy (XPS) analysis using an AXIS-ULTRA DLD (Shimadzu/Kratos) with an Al-Kα X-ray radiation ($hv$ = 1486.6 eV), operating at a pressure of $9 \times 10^{-9}$ Torr. The analyzed area was approximately $1 \times 1$ mm$^2$. Electrical conductivity was measured using a Physical Properties Measurements System (PPMS, Quantum Design) in the temperature range of 2–390 K.

XPS analysis was carried out to examine the chemical states of raw Sn in the both grain (red line) and powder (blue line) forms (Fig.1). There is a well-defined peak at 487.5 eV in the XPS spectrum of the Sn 3d, which is associated with the Sn $3d_{5/2}$ state. This is a characteristic feature of a tetravalent Sn ion (Sn$^{4+}$) [20], implying the presence of SnO$_2$ in both grain and powder Sn. Most likely, it could be on the surfaces. The smaller peak at around 485 eV arises from the Sn$^0$ state [20], indicating the presence of the non-oxidized metal Sn. The thickness of SnO$_2$ layer should be less than few nm that is the value of the penetration depth of XPS measurement. These results indicate that by using the powder Sn the larger amount of oxygen contamination occurs during the synthesis due to larger volume fraction of SnO$_2$ coming from the larger surface area.

Figure 2 shows the X-ray diffraction patterns of pulverized SnSe single crystals grown from grain and powder Sn. Inset shows the enlargement of the (400) reflection. The diffraction patterns of both samples can be indexed with an orthorhombic structure of the space group *Pnma*. The lattice constants of *a*, *b* and *c* axes were determined to be *a*

= 11.48(7) Å, $b$ = 4.44(2) Å and $c$ = 4.15(1) Å for both samples comparable to the literature.

Figure 3 shows the temperature dependence of the electrical conductivity of SnSe single crystals along the $bc$ plane. The electrical conductivity depends strongly on the ratio of grain and powder form Sn used as starting materials. The sample from grain Sn 100%, which may contain the smallest oxygen contamination amount, showed semiconducting behavior. The conductivity sharply decreased with decreasing temperature. The small electrical conductivity indicates a low hole concentration ensuring the purity of the sample that has small deviation from the ideal stoichiometry. On the other hand, the electrical conductivities increased and their behaviors became metal like with decreasing the ratio of grain Sn. The electrical conductivity of SnSe grown by 100% powder Sn showed good agreement with that of the reported ones by Zhao *et al*. [1]. The present results indicate that transport properties of SnSe single crystal can be controlled from semiconductor to metal by varying the ratio of grain and powder form of Sn used as a starting material. This can be due to the difference of the oxygen volume fraction presenting on the surface of both Sn.

We found that the different electrical conductivity reported by several groups could be intrinsically related to the contamination of oxygen during the synthesis. We succeeded in controlling the transport properties of SnSe single crystals by changing the oxygen amount presented in raw Sn as a surface oxidization. The electrical conducting behavior varied from semiconductor to metal with increase of oxygen amount.

This work was partly supported by JST CREST Grant No. JPMJCR16Q6, and JSPS KAKENHI Grant No. JP16J05432.

‡Corresponding author: Aichi Yamashita

E-mail: YAMASHITA.Aichi@nims.go.jp

Postal address: National Institute for Materials Science, 1-2-1 Sengen, Tsukuba, Ibaraki 305-0047, Japan

Tel.: (+81)29-851-3354 ext. 2976

[1] Li-Dong Zhao, Shih-Han Lo, Yongsheng Zhang, Hui Sun, Gangjian Tan, Ctirad Uher, C. Wolverton, Vinayak P. Dravid & Mercouri G. Kanatzidis, Nature **508**, 373 (2014)

[2] Yajie Fu, Jingtao Xu, Guo-Qiang Liu, Jingkai Yang, Xiaojian Tan, Zhu Liu, Haiming Qin, Hezhu Shao, Haochuan Jiang, Bo Liang, and Jun Jiang, J. Mater. Chem. C, **4**, 1201 (2016)

[3] J. C. Li, D. Li, X.Y. Qina, J. Zhang, Scripta Materialia **126**, 6–10, (2017)

[4] Yi Li, Bin He, Joseph P. Heremans, Ji-Cheng Zhao, Journal of Alloys and Compounds **669**, 224 (2016)

[5] Yulong Li, Xun Shi, Dudi Ren, Jikun Chen, and Lidong Chen, Energies, **8**, 6275 (2015)

[6] Xue Wang, Jingtao Xu, Guoqiang Liu, Yajie Fu, Zhu Liu, Xiaojian Tan, Hezhu Shao,

Haochuan Jiang, Tianya Tan , and Jun Jiang, Appl. Phys. Lett. **108**, 083902 (2016)

[7] Guodong Tang, Qiang Wen, Teng Yang, Yang Cao, Wei Wei, Zhihe Wang, Zhidong Zhangb and Yusheng Li, RSC Adv., **7**, 8258 (2017)

[8] Eyob K. Chere, Qian Zhang, Keshab Dahal, Feng Cao, Jun Mao, and Zhifeng Ren, J. Mater. Chem. A, **4**, 1848(2016)

[9] Qian Zhang, Eyob Kebede Chere, Jingying Sun, Feng Cao, Keshab Dahal, Shuo Chen, Gang Chen, and Zhifeng Ren, Adv. Energy Mater., **2015**, 1500360 (2015)

[10] Hongliang Liu, Xin Zhang, Songhao Li, Ziqun Zhou, Yanqin Liu, and Jiuxing Zhang, Journal of Electronic Materials, **46**, (2017)

[11] M. Gharsallah, F. Serrano-Sánchez, N. M. Nemes, F. J. Mompeán, J. L. Martínez, M. T. Fernández-Díaz, F. Elhalouani & J. A. Alonso, Scientific Reports **6**, 26774 (2016)

[12] Anh Tuan Duong, Van Quang Nguyen, Ganbat Duvjir, Van Thiet Duong, Suyong Kwon, Jae Yong Song, Jae Ki Lee, Ji Eun Lee, SuDong Park, Taewon Min, Jaekwang Lee, Jungdae Kim & Sunglae Cho, Nature Communications **7**, 13713 (2016)

[13] Kunling Peng, Xu Lu, Heng Zhan, Si Hui, Xiaodan Tang, Guiwen Wang, Jiyan Dai, Ctirad Uher, Guoyu Wang, Xiaoyuan Zhou, Energy and Environmental Science, **9**,

454-460 (2016)

[14] Cheng-Lung Chen, Heng Wang, Yang-Yuan Chen, Tristan Day, G. Jeffrey Snyder, J. Mater. Chem. A, **2**, 11171-11176, (2014)

[15] D. Ibrahim, J.-B. Vaney, S. Sassi, C. Candolfi, V. Ohorodniichuk, P. Levinsky, C. Semprimoschnig, A. Dauscher, and B. Lenoir, Appl. Phys. Lett. **110**, 032103 (2017)

[16] F. Serrano-Sánchez, M. Gharsallah, N. M. Nemes, F. J. Mompean, J. L. Martínez, and J. A. Alonso, Appl. Phys. Lett. **106**, 083902 (2015)

[17] Niraj Kumar Singh, Sivaiah Bathula, Bhasker Gahtori, Kriti Tyagi, D. Haranath, Ajay Dhar, Journal of Alloys and Compounds, **668**, 152-158, (2016)

[18] S. Sassi, C. Candolfi, J.-B. Vaney, V. Ohorodniichuk, P. Masschelein, A. Dauscher, and B. Lenoir, Appl. Phys. Lett. **104**, 212105 (2014)

[19] Tian-Ran Wei, Chao-Feng Wu, Xiaozhi Zhang, Qing Tan, Li Sun, Yu Pan and Jing-Feng Li, Phys. Chem. Chem. Phys., **17**, 30102-30109, (2015)

[20] L. Kövér, Zs. Kovács, R. Sanjinés, G. Moretti, I. Cserny, G. Margaritondo, J. Pálinkás, H. Adachi, Surface and Interface Analysis, **23**, 461-466, (1995)

Fig. 1 Sn $3d_{5/2}$ XPS spectrum for the surface of grain and powder Sn.

Fig. 2 XRD patterns of the pulverized SnSe single crystals synthesized from grain and powder Sn.

Fig. 3 Temperature dependence of the electrical conductivity of SnSe single crystals along the *bc* plane grown by using different ratio of grain and powder form Sn as a starting material.

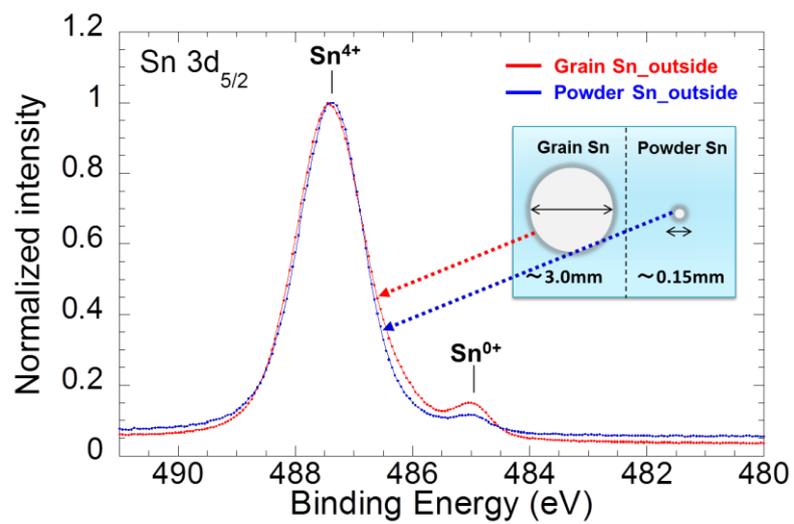

Fig.1

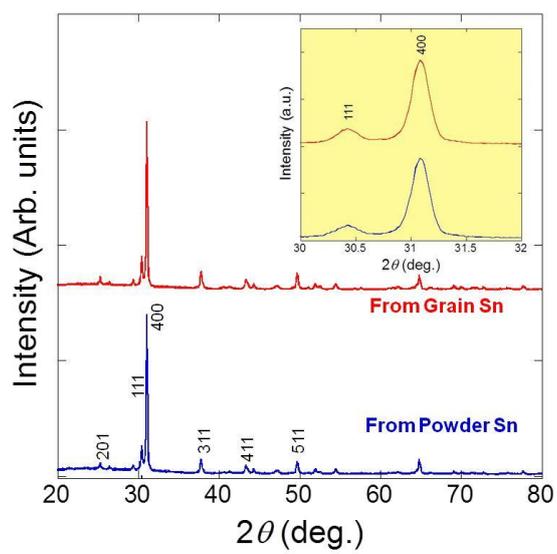

Fig.2

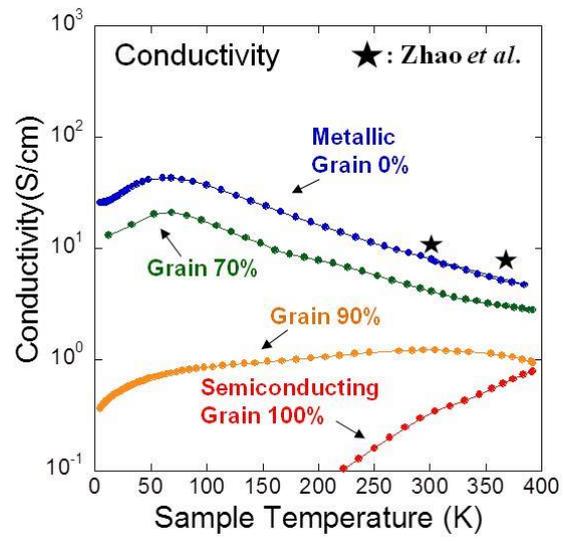

Fig.3